\documentclass[twoside]{article}

\usepackage{graphicx}%
\usepackage{multirow}%
\usepackage{amsmath,amssymb,amsfonts}%
\usepackage{amsthm}%
\usepackage{mathrsfs}%
\usepackage[title]{appendix}%
\usepackage{xcolor}%
\usepackage{textcomp}%
\usepackage{manyfoot}%
\usepackage{booktabs}%
\usepackage{algorithm}%
\usepackage{algorithmicx}%
\usepackage{algpseudocode}%
\usepackage{listings}%
\usepackage[inline]{enumitem}
\usepackage{xurl}
\usepackage[preprint]{aistats2026}
\usepackage[round]{natbib}

\begin{document}

%

%
\runningauthor{Paolo Giudici, Rosa C. Rosciano, Johanna Schrader, Delf-Magnus Kummerfeld}

\twocolumn[

\aistatstitle{A Multiclass ROC Curve}

\aistatsauthor{ Paolo Giudici \\ Department of Economics,\\
University of Pavia,\\ Pavia, Italy \And Rosa C. Rosciano \\ Department of Economics,\\ University of Pavia,\\ Pavia, Italy 
\AND
Johanna Schrader \\ L3S Research Center,\\ CAIMed  - Lower Saxony Center for AI\\ \& Causal Methods in Medicine,\\Hannover, Germany \And Delf-Magnus Kummerfeld \\ Leibniz Institute on Aging,\\ Fritz Lipmann Institute (FLI),\\ Jena, Germany}

\aistatsaddress{ } ]

\begin{abstract}
This paper introduces a novel methodology for constructing multiclass ROC curves using the multidimensional Gini index. 
The proposed methodology leverages the established relationship between the Gini coefficient and the ROC Curve and extends it to multiclass settings through the multidimensional Gini index. 
The framework is validated by means of two comprehensive case studies in health care and finance. 
The paper provides a theoretically grounded solution to multiclass performance evaluation, particularly valuable for imbalanced datasets, for which a prudential assessment should take precedence over class frequency considerations.
\end{abstract}

\section{INTRODUCTION}
The aim of this work is to develop a novel methodology to extend the well-known Receiver Operating Characteristics (ROC) curves from binary to multiclass response classification problems, such as credit rating assessments and cancer type identification. 


The multiclass nature of the response introduces significant challenges.
The most important issue concerns the choice of thresholds on which to base rating assessments. While in binary classification, one threshold is sufficient, multiclass classification requires many of them, increasing the subjectivity of the assessments.

Another important problem is class imbalance, which can severely distort traditional performance metrics based on classification errors, precision, and recall. With imbalanced data, models can exhibit high overall accuracy while demonstrating failure in minority classes. 


The above problems have been overcome in the binary case with the introduction of the Receiver Operating Characteristic (ROC) curve~\citep{Fawcett}, a cornerstone of binary classification that plots the True Positive Rate against the False Positive Rate across all decision thresholds. The ROC curve can solve the first problem, as it does not depend on the chosen threshold; and it can solve the second problem, as it can be employed only for the regions of interest, leading to the Partial ROC Curve~\citep{Walter2005}.

In addition, the ROC curve not only provides a visual representation, but also a single summary statistic, the Area Under the ROC curve (AUC), which represents an easy to interpret summary measure,  normalized in $[0,1]$. The AUC is one of the most employed measures of predictive performance in classification problems for binary responses, similarly to the mean squared error in predictive problems for continuous responses.

Against the above advantages, the ROC Curve lacks a universally accepted counterpart for multiclass responses. Some attempts have been made, using a One-vs-One or One-vs-Rest comparison~\citep{HandT}; however, none of them has reached enough generality and practical implementability.
The One-versus-Rest (OvR) and One-versus-One (OvO) approaches to multiclass ROC and AUC analysis suffer from fundamental limitations. OvR approaches create severe class imbalance when treating single classes against all others, while OvO methods generate multiple pairwise comparisons that become difficult to interpret when the number of classes increases. Furthermore, neither of these methods provides a single ROC curve that can be easily interpreted and validated~\citep{HandT,Landgrebe}.
Other attempts do not have sufficient generality. For example, the volume under the ROC of~\cite{Zhou2002} applies only to ternary classifications. 

In this paper, we address this gap by proposing a novel methodology that constructs a single unified ROC curve for multiclass classification tasks using the multidimensional Gini index of~\cite{Giudici2024}. Our approach leverages the established relationship between the Gini coefficient and ROC analysis in binary classification, extending it to multiclass settings, thereby preserving the intuitive decision-theoretic properties of the  ROC Curve.

Our main contribution is to develop a theoretically grounded approach to construct a single ROC curve for multiclass classification, based on a robust statistical metric, the Gini index, whose extension to the multidimensional case also provides class-specific weights for the multidimensional ROC Curve.

From the applied viewpoint, we apply the proposed methodology to credit rating assessments, a very relevant application problem, in which the response is multiclass and ordinal, often approximated as binary for the absence of a general evaluation tool. In a second health care use-case, we apply the method to a cancer type prediction task using genetic data. Here, the multiclass data is nominal and severely imbalanced.


\section{LITERATURE REVIEW}

While the binary ROC framework provides a well-established methodology for performance assessment, its extension to multiclass problems introduces significant conceptual and practical challenges. To the best of our knowledge, there exists no single universally accepted measurement tool that comprehensively summarizes model predictive performance in the multiclass context, see~\citep{HandT}.

Indeed, several approaches have been proposed to extend ROC analysis beyond binary classification. The most common include the One-versus-Rest (OvR) approach, the One-versus-One (OvO) approach, and the Volume Under the Surface (VUS) methodology.
    
The One-versus-Rest approach decomposes the multiclass problem into multiple binary classification problems, constructing one ROC curve for each class treated as the positive class against all others combined. This generates \textit{k} different ROC curves for a \textit{k}-class problem~\citep{Provost}.
A multiclass AUC measure can then be calculated as a simple average of the single AUC.
    
While straightforward to implement, this method suffers from several limitations.
First, when treating a single class as positive against all others, severe class imbalance often arises, particularly for rare classes, potentially biasing the resulting ROC curves~\citep{Fawcett}.
Second, aggregating all non-target classes masks the model's ability to discriminate between specific class pairs, potentially obscuring important performance characteristics~\citep{Landgrebe}. For instance, misclassifying class A as class B might have different practical implications than misclassifying class A as class C. This nuance is difficult to capture in aggregated performance metrics.
Third, different thresholds may be optimal for different class-specific ROC curves, making it difficult to select a coherent set of thresholds for the overall multiclass classifier, as argued in~\cite{Ferri}.

The One-versus-One approach constructs ROC curves for each possible pair of classes, generating $(k(k-1))/2$ curves for a \textit{k}-class problem. \cite{HandT} proposed a multiclass AUC measure based on averaging all pairwise AUCs, known as the M-measure

    \[
    M = \frac{2}{k(k-1)} \sum_{i < j} A(i, j),
    \]

where $A(i, j)$ represents the AUC for discriminating between classes $i$ and $j$. 
    
While this approach captures pairwise separability between classes, it presents several drawbacks. First, analyzing and interpreting multiple pairwise ROC curves becomes challenging, complicating the derivation of actionable insights~\citep{Flach}.
Second, the pairwise approach treats all class pairs equally, potentially overlooking naturally ordered or hierarchical relationships between classes~\citep{Ferri}.
Third, the M-measure's simple averaging of pairwise AUCs fails to account for the varying importance of different class distinctions and can mask severe performance deficiencies in distinguishing particular class pairs~\citep{Hand}.

While both OvR and OvO reduce the multiclass problem into multiple binary problems, the Volume Under the Surface method~\citep{Zhou2002} directly addresses a multiclass response.
Specifically, it has been proposed to extend the ROC curve to a three-dimensional ROC surface, with the Volume Under the Surface (VUS) serving as the multiclass analog of AUC~\citep{Mossman}. This approach can be generalized, in principle, to \textit{k} dimensions for \textit{k} classes, but faces severe limitations.
First, beyond three classes, the resulting hypersurface cannot be directly visualized, severely limiting intuitive interpretation~\citep{Landgrebe}.
Second, the probabilistic interpretation of VUS becomes increasingly complex and less intuitive, compared to the binary AUC~\citep{HandT}.

The limitations of the previous approaches have been tackled by means of various weighted averaging schemes for the AUC, which combine class-specific or pairwise information, adjusting for factors such as class prevalence or misclassification costs~\cite{Ferri}. However, these approaches face other problems. First, determining appropriate weights remains subjective and context-dependent, potentially introducing bias~\citep{Landgrebe}.
Second, weighted averages may obscure performance deficiencies for specific classes or class combinations~\citep{Flach}. Third, identifying optimal operating points across multiple weighted ROC curves remains problematic~\citep{Ferri}. A final relevant problem is that most weighted schemes lack a coherent theoretical foundation, making their interpretation and justification difficult~\citep{Hand}.

The above-mentioned limitations of existing multiclass ROC extensions highlight the need for a more comprehensive framework that preserves the intuitive interpretation and decision-theoretic foundations of the ROC curve in the binary case while accommodating the complexity of multiple response classification. 
We aim to fill this gap, proposing a multiclass ROC curve that leverages the properties of a recently proposed multidimensional Gini index to derive an endogenous and consistent system of weights for binary ROC curves that converts them into a multiclass ROC curve and a related Area Under the Curve.

\section{PROPOSAL}

Before introducing our proposal, let us briefly recall the Gini index and its connection with the ROC Curve.

The Gini index has its main application in the study of economic inequality, through the Lorenz curve~\citep{Lorenz}, which illustrates the cumulative proportions of wealth. The Gini coefficient measures inequality as twice the area between the Lorenz curve and the diagonal (line of equality)~\citep{DmitrievDobriban2022,Schechtman}. Formally:

\[
G = 1 - 2\int_0^1 L(F)\, dF,
\]

where \(L(F)\) is the Lorenz curve, that is, the cumulative share of total outcomes up to fraction \(F\) of the population.
In the binary case, the Lorenz curve is a statistical representation of the Cumulative Distribution Function (CDF) of the positive cases of the target class and is related to the Area Under the ROC Curve (AUC). 

The Gini index, originally developed to measure income inequality in economics, has a well-established relationship with ROC analysis in the binary classification context. The Gini index $G$ can be expressed as
\begin{equation} AUC = \frac{G+1}{2},
\label{Gini}
\end{equation} 
\citep{HandT}.

The relationship in (\ref{Gini}) provides an alternative perspective on classifier performance, measuring the relative degree of separation of a distribution, as the area between the ROC curve and the ROC curve of a baseline random model~\citep{Flach}. 

The  AUC and the Gini index are normalized in $[0,1]$, and are inversely related: while $AUC=1$ indicates the ideal case of perfect classification, and $AUC=0$ a random classification, $G=0$ indicates maximum income equality and $G=1$ maximum income inequality. 
We will exploit this relationship to construct an aggregate multiclass ROC curve that preserves the interpretability of the binary case while addressing the limitations of existing multiclass approaches.
More precisely, we will assume that the relationship in (\ref{Gini}) holds when the AUC is calculated for a multiclass response, as follows  

\begin{equation} AUC_{agg} = \frac{G_{agg}+1}{2},
\label{Gini2}
\end{equation} 

where $G_{agg}$ is a multidimensional Gini index.  

Equation (\ref{Gini2}) moves the research problem from the proposition of a multidimensional ROC to the proposition of a multidimensional Gini index. 

Building a multidimensional Gini index poses substantial difficulties. Univariate inequality measures such as the Lorenz curve and the Gini index~\citep{Gini14,Gini21} are foundational tools for analyzing the mutual variability expressed by statistical distributions, and  their properties should be maintained when extending them to multivariate data. 

In particular, an important property that any multivariate Gini index should possess is scale invariance, which ensures that the index remains unchanged when components of the original random vector are scaled by positive constants~\citep{Hurley}. 
This characteristic is crucial for assessing inequality within multivariate scenarios that may involve differing units of measurement. This situation arises, for instance, when we employ the multidimensional Gini index to evaluate the dispersion of the predicted probabilities by a classifier across various classes in the multiclass setting.

\cite{Giudici2024} proposed a multivariate Gini index that satisfies the scale invariance property by utilizing the Mahalanobis distance in place of the Euclidean distance. 
While the Euclidean distance is commonly used to measure similarity or dispersion in multivariate data, it does not account for differences in scale or correlations between variables.
The Mahalanobis distance, on the other hand, achieves scale invariance by normalizing for variances and covariances. The Mahalanobis distance transforms the data such that each feature has unit variance and all features are uncorrelated. As a result, the transformed matrix is of full rank, ensuring that its feature columns are linearly independent. This adjustment allows the Mahalanobis distance to accurately capture true statistical dispersion, regardless of scale or inter-variable relationships.

Leveraging the Mahalanobis distance, \cite{Giudici2024} defined a multivariate Gini as a convex combination of one-dimensional Gini indices.
Their result is based on the notion of whitening.
Whitening is a linear transformation that converts a random \textit{n}-dimensional vector $\mathbf{X} = (X_1, \ldots, X_n)^T$ with mean value $\mathbf{m} = (m_1, \ldots, m_n)^T$ and positive definite $n {\times} n$ covariance matrix $\Sigma$ into a new random vector $\mathbf{X}^*=W_\mu X$ whose entries are orthonormal, meaning that the variance of each $X_i^*$ is 1 and the covariance of any $X_i^*$ and $X_j^*$ is zero whenever $i \neq j$~\citep{Giudici2024}. The transformation can be viewed as a generalization of standardization, which is carried out by

\begin{equation}
\mathbf{X}^{'} = V^{-1/2} \mathbf{X},
\end{equation}

where the diagonal matrix $V = \text{diag}(\text{var}(X_1), \text{var}(X_2), \ldots, \text{var}(X_n))$ contains the variances of $X_i$. While standardization ensures unitary variance for each component, it does not remove the correlations between components~\citep{Kessy}. 

In its most generic form, a whitening process is a map

\begin{equation}
S:P_{\Sigma}^n (\mathbb{R}^n) \rightarrow P_{Id}(\mathbb{R}^n),
\end{equation}

that transforms an $n$-dimensional random vector $\mathbf{X}$ characterized by a probability measure $\mu$, with positive mean $\mathbf{m}$ and covariance matrix $\Sigma$, into a new $n$-dimensional random vector whose covariance matrix is the identity matrix~\citep{Giudici2024}.

A whitening process $S$ is linear if, for any given $\mathbf{X} \in P(\mathbb{R}^n)$, there exists an $n \times n$ square matrix that depends on $\mathbf{X}$ through its distribution, denoted $W_\mu$, such that

\begin{equation}
\mathbf{X}^* = S(\mathbf{X}) = W_\mu \mathbf{X}.
\end{equation}

The matrix $W_\mu$ is known as the whitening matrix associated with $S$~\citep{Kessy}. If the covariance matrix $\Sigma$ of $\mathbf{X}$ is invertible, the whitening matrix satisfies
\begin{equation}
W_\mu \Sigma W_\mu^T = I,
\end{equation}
which simplifies to
\begin{equation}
W_\mu^T W_\mu = \Sigma^{-1}.
\end{equation}
This condition does not uniquely determine the linear application that transforms $\mathbf{X}$ into a whitened vector, allowing for rotational freedom. 

Based on the properties of scale invariance and non-negativity preservation, \cite{Giudici2024} recommend using
the correlation whitening, also known as Zero-Components Analysis ($ZCA-cor$). The $ZCA-cor$   whitening matrix is defined as
\begin{equation}
W_\mu^{ZCA-cor} = P^{-1/2} V^{-1/2},
\end{equation}
where $P$ is the correlation matrix of $\mathbf{X}$, and $V$ is the diagonal matrix containing the variances of each component~\citep{Giudici2024}.

This transformation allows preserving the range [0,1] when applied to non-negative data, consistent with the univariate Gini index, and it enables the construction of a multivariate Gini index that maintains desirable properties while accounting for the correlation structure among variables.
In addition, ZCA whitening has the advantage of preserving the orientation of the data in the original space, making it useful for applications where interpretability is important~\citep{Kessy}, as in our credit rating and health care use-cases.

\cite{Giudici2024} proposed a multidimensional Gini index,  $G_1(X)$,  which can be expressed as a convex combination of the one-dimensional Gini indices of the components of the vector $W_\mu^{ZCA}X$

\begin{equation}
G_1(X) =
\sum_{i=1}^n
\frac{|m_i^*|}{\sum_{j=1}^n |m_j^*|} \cdot
G\left( (W_\mu^{\mathrm{ZCA}} X)_i \right),
\label{eq:multigini}
\end{equation}

where $m$ is the mean of $X$, $m_i^{*}=(W_{\mu}^{ZCA}m)_i$ and $G(W_{\mu}^{ZCA}X)_i$ is the one-dimensional Gini index of the $i-th$ component of $W_\mu^{ZCA}X$.
This equation is particularly significant as it establishes that the higher-dimensional Gini index induced by the  $l_1$-Mahalanobis norm is a convex combination of the one-dimensional Gini indices of the whitened random variables. 
It is the pivotal result on which to construct the Multiclass ROC curve. 

Among other properties, it can be shown that, when the components of $X$ are non-negative, the index satisfies
$$0 \leq G_1(X) \leq 1.$$
This property ensures that the multidimensional Gini index remains within the same interpretable range as its univariate counterpart, thereby justifying equation~\eqref{Gini2}.

The Multidimensional Gini index offers a coherent way to measure inequality in multivariate data by leveraging whitening transformations to express the same inequality as a function of one-dimensional Gini indices applied to whitened components. 
The weights in the convex combination depend on the relative importance of the mean of each whitened component, normalized by the sum of all mean values:

\begin{equation}
w_i = \frac{|m_i^*|}{\sum_{j=1}^{n} |m_j^*|}.
\label{eq:weights}
\end{equation}

These weights will be used to construct the Multiclass Gini Multidimensional ROC curve. 
In the application sections, we will illustrate how a multiclass ROC curve can be computed using the weights obtained from the multidimensional Gini index applied to the ZCA whitened predictions. We will also compare different machine learning models using the resulting Area Under the multiclass ROC Curve. 

The obtained multiclass AUC will be compared with that obtained using micro and macro AUC approaches, 
 computed by constructing individual class-wise ROC curves through the One-vs-Rest approach. The difference is that, while macro AUC is the unweighted mean of the per-class AUCs, treating each class equally,  micro AUC is obtained by a weighted mean that gives higher weights to the most accurate classes.





\section{APPLICATION TO FINANCE}

Our first case study employs a panel dataset provided by Modefinance, an italian credit rating agency, which includes financial, credit risk, and Environmental, Social, and Governance (ESG) metrics for 1,803 Small and Medium-sized Enterprises (SMEs). The dataset spans three consecutive years (2020-2022) and comprises 38 variables, among which is credit rating, the response variable, a set of variables from the balance sheets of the companies, the ESG scores, and the country and economic sector of the companies. More details on the data can be found in \cite{Chen2024}.

\begin{figure}[ht]
    \centering
    \includegraphics[width=0.9\linewidth]{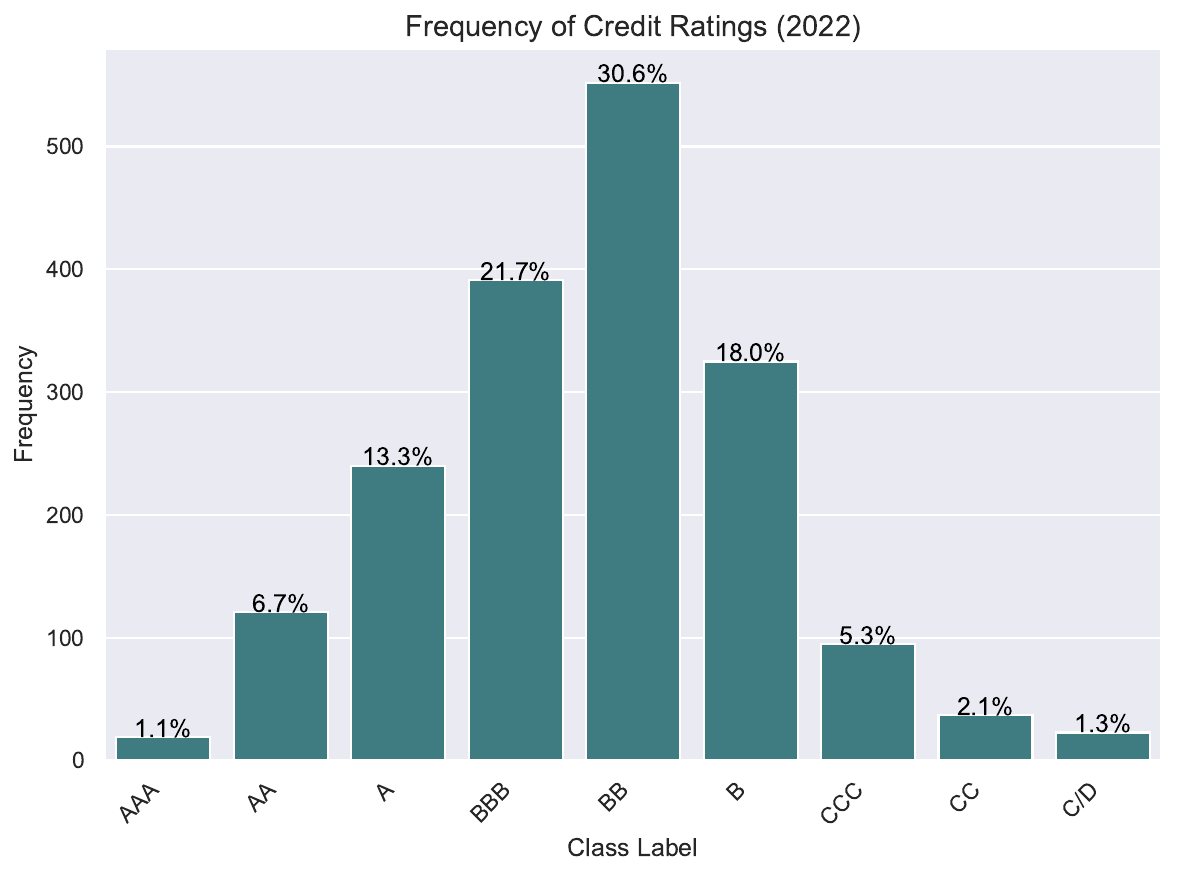}
    \caption{Frequency of 2022 credit-rating classes}
    \label{fig:class_dist}
\end{figure}

As illustrated in Figure~\ref{fig:class_dist}, the frequency distribution shows a concentration of observations within intermediate rating categories. The distribution exhibits a clear variability in class size, with very few companies appearing at the extremes (AAA, CC, C\&D). This skewed pattern reflects the typical risk profile of established SMEs, where exceptionally strong credit standings and defaults are relatively rare. 
Overall, the dataset presents a particularly demanding environment for classification: a relatively limited sample size, strong multicollinearity among financial variables, and severe class imbalance. These factors collectively constrain the achievable performance of classification algorithms. However, this creates an excellent opportunity to test how our multiclass ROC curve behaves under challenging real-world conditions.

We have applied four alternative machine learning models to the previously described data, using the same cross-validation sample, with 80\% of the observations in the training set and 20\% in the test set.
All models aim to predict a 9-dimensional vector of binary variables, each of which indicates whether the company considered belongs to the respective rating class or not.
For each model, we obtained the multivariate predictions for the observations in the test set and then applied the ZCA-correlation whitening matrix to the predicted probabilities. Note that, as the binary variables are orthogonal by design, the whitening transformation essentially scales all variances to one, preserving rankings, which is essential for the ROC analysis.


We then compute the multidimensional Gini coefficient from the whitened predictions using Equation~\eqref{eq:multigini}, obtaining the class-specific weights in Equation~\eqref{eq:weights}. As the binary response variables are not correlated, the weights are obtained as ratios between the estimated mean and standard deviation of binomial distributions. This implies that higher weights are assigned to classes with a higher probability of success and a higher imbalance.

\begin{figure}[htbp]
\centering
\includegraphics[width=0.9\linewidth]{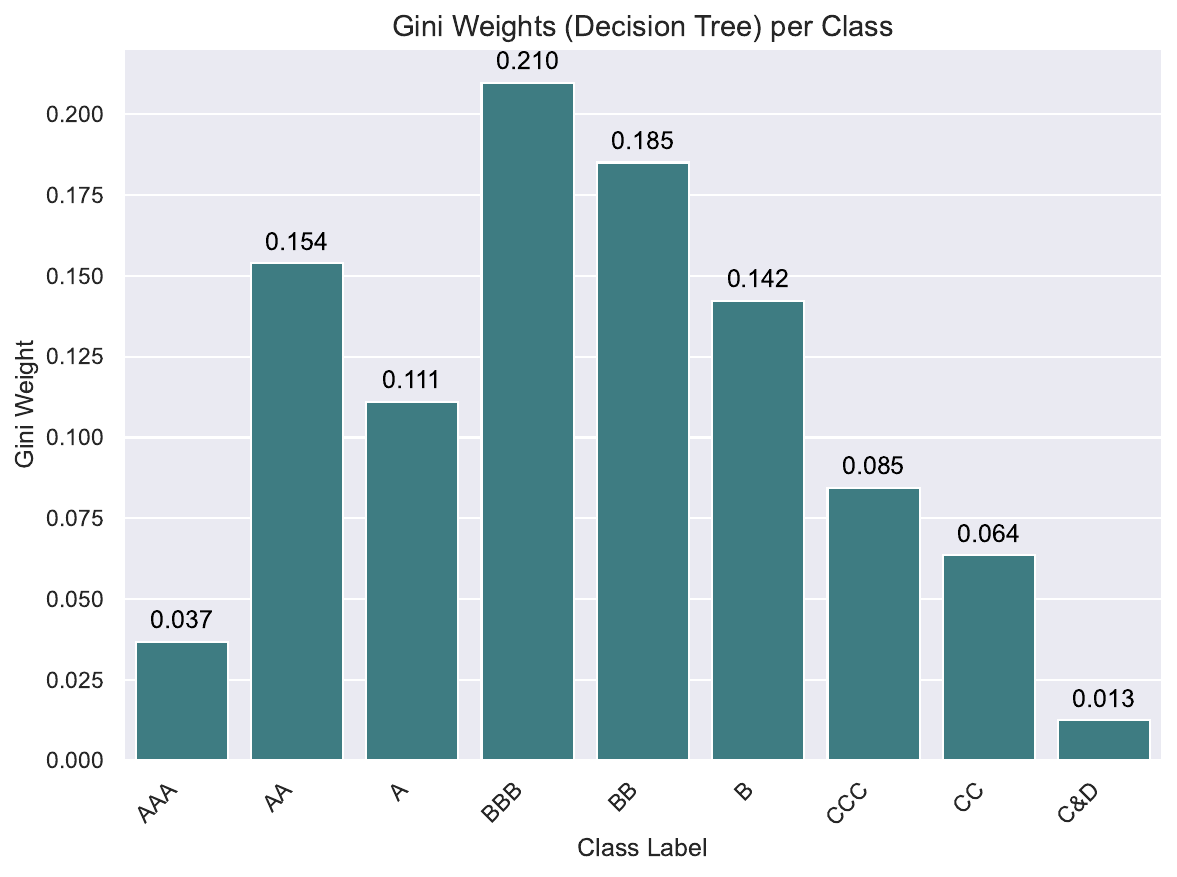}
\caption{Gini weights for a Decision Tree classifier showing relative class contributions to the aggregate ROC curve on the finance dataset.}
\label{fig:weights}
\end{figure}

To exemplify, Figure~\ref{fig:weights} illustrates the weight distribution across credit rating classes for the decision tree model. 
Class BBB receives the highest weight (0.21),  while class C \& D receives the lowest weight (0.013). Note that class BB, the most frequent in  in Figure~\ref{fig:class_dist}, receives a weight equal to 0.185, lower than that of class BBB. This is the result of the correction for unbalancedness operated by the Gini weights: BB is less frequent and, thus, more unbalanced than BBB. Figure~\ref{fig:class_dist} also indicates that, among rare classes, AAA is  more unbalanced than C\&D. The unbalancedness correction takes this into account and produces a Gini weight higher for AAA than for C\&D, as can be seen in 
Figure~\ref{fig:weights}.

We recall that, in macro-averaging, the AUCs are equally weighted, whereas, in micro-averaging, the most accurate classes receive higher weights. Differently, in the Gini-based multiclass AUC, the classes with higher unbalancedness receive higher weights, as can be seen comparing Figure~\ref{fig:class_dist} with Figure~\ref{fig:weights}.

\begin{figure}[htbp]
\centering
\includegraphics[width=0.9\linewidth]{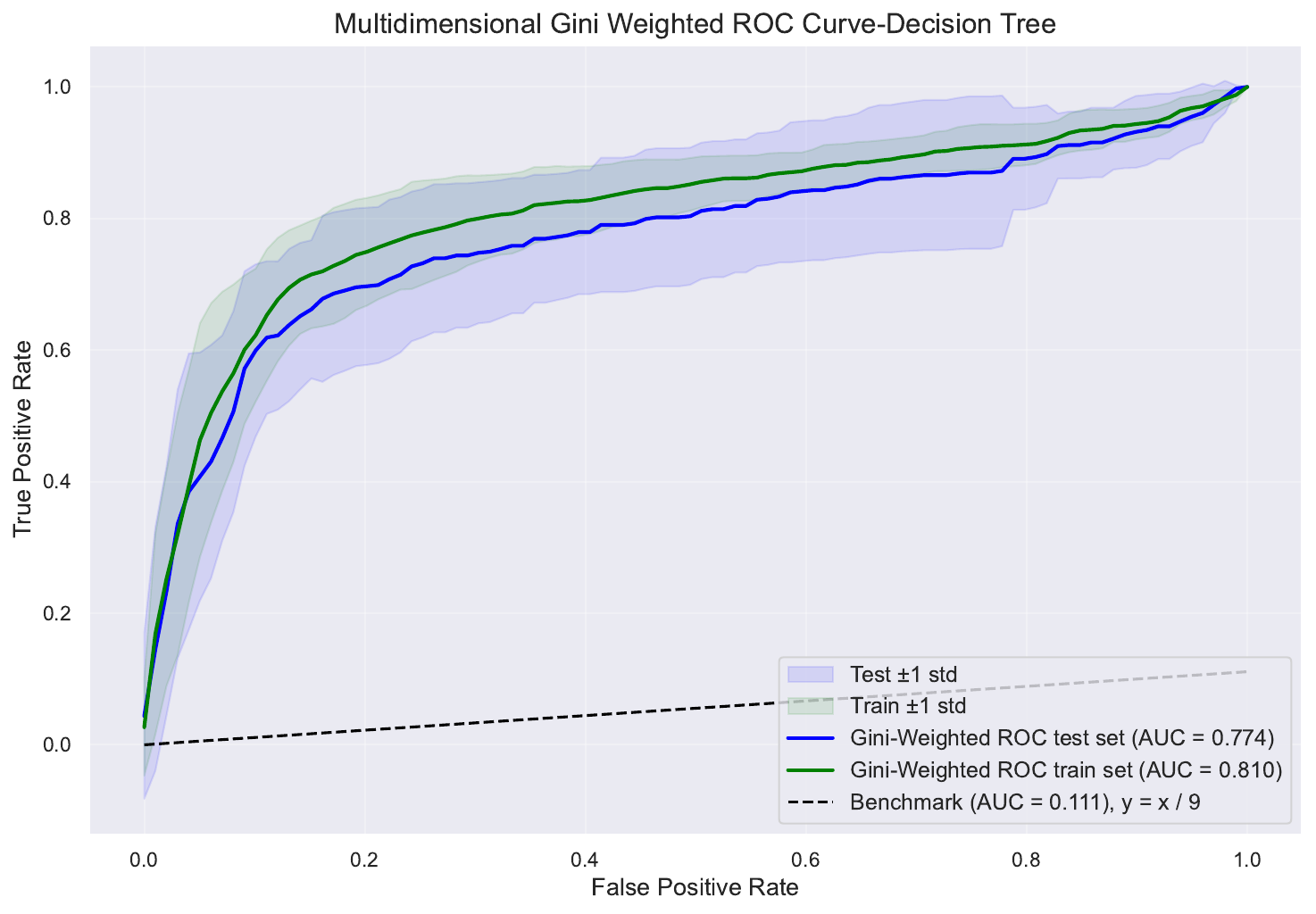}
\caption{Multiclass ROC curve for a Decision Tree classifier on the finance dataset.}
\label{fig:multiclass-roc-rf}
\end{figure}

The same weights are used to calculate aggregate True Positive Rates and False Positive Rates, for a set of given thresholds, thus leading to a multiclass ROC curve that can be compared with the individual ones. In addition, by means of resampling, we can calculate several multiclass ROC curves, thereby providing standard errors and confidence bounds.
Figure \ref{fig:multiclass-roc-rf} shows the multiclass ROC curve, with the related confidence bounds, for the decision tree model. In the Figure, we report the calculation of the curve also for the training set, along with the random baseline (corresponding to an AUC of 1/9).
As expected, the model performs better on the training set (in-sample predictions) than on the test set (out-of-sample predictions).

\begin{figure}[htbp]
\centering
\includegraphics[width=\linewidth]{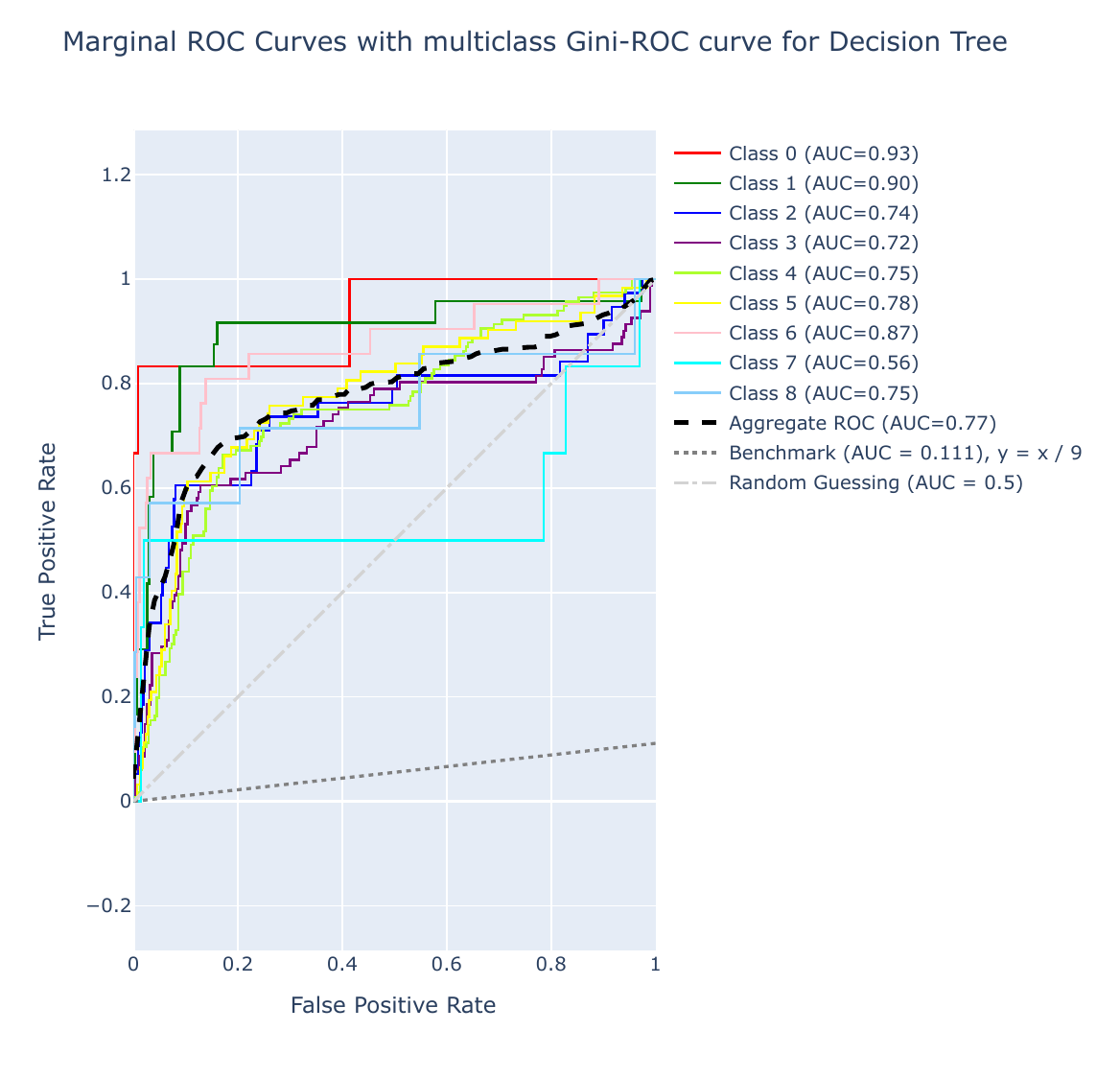}
\caption{Individual ROC curves and multiclass ROC curve for the Decision Tree model on the finance dataset.}
\label{fig:per-class-multiclass-roc}
\end{figure}

Figure~\ref{fig:per-class-multiclass-roc} compares the multiclass ROC curve with the individual ROC curves, one for each class, in order of rating. As expected, the aggregate curve lies between the individual class curves. We can also observe that class 7 (corresponding to rating CC) is the most difficult to predict. It is indeed highly unbalanced. 

\begin{table*}[ht]
\caption{Comparison of AUC Metrics Across Different Classifiers }
\label{tab:classifier_metrics_comparison_reduced}
\begin{center}
\begin{tabular}{lcccccc}
\textbf{CLASSIFIER} & \multicolumn{2}{c}{\textbf{GINI AUC }} & \multicolumn{2}{c}{\textbf{MACRO AUC}} & \multicolumn{2}{c}{\textbf{MICRO AUC}} \\
& Finance & Genetic & Finance & Genetic & Finance & Genetic \\
\hline
Decision Tree & 0.77 & 0.79 & 0.80 & 0.87 & 0.85 & 0.94 \\
Logistic Regression & 0.51 & 0.94 & 0.80 & 0.98 & 0.88 & 0.99 \\
Random Forest & 0.67 & 0.67 & 0.89 & 0.94 & 0.90 & 0.96 \\
Bagging Classifier & 0.44 & 0.95 & 0.83 & 0.98 & 0.86 & 0.99 \\
\end{tabular}
\end{center}
\end{table*}

Table \ref{tab:classifier_metrics_comparison_reduced} summarizes the comparison of the machine learning models in terms of the proposed multiclass AUC. 
The decision tree model emerges as the best classifier for the finance task, achieving a Gini  AUC of 0.77, followed by the Random Forest model. Micro and Micro AUC give similar values to all models, with the Random Forest model slightly more accurate.  Overall,  the Gini AUC seems to better discriminate among different models, with a larger range of values. 
This is because the proposed methodology fundamentally differs from traditional AUC averaging, as it weights classes according to their unbalancedness and frequency, rather than equally, or by giving higher weight to the most accurate classes. 
In other words, the Gini-based approach appears more conservative and prudential than macro averaging and micro averaging.
Such a conservative assessment is indeed advantageous for credit risk applications, as it reduces the risk of deploying overconfident models in high-stakes financial decisions. 
 The comparison with other multiclass methods could, however, serve as a quality indicator,  in which agreement indicates robustness.

\section{APPLICATION TO HEALTH CARE}
The second case study uses a clinical dataset of 16,188 samples with a feature space consisting of expression data from 2,565 genes~\citep{matsuzaki2023prediction}. Each sample belongs to one of 19 disease classes (5 benign non-cancerous diseases, 13 malignant cancers, and one healthy class). The dataset from~\cite{matsuzaki2023prediction} is publicly available on the Gene Expression Omnibus (GEO) repository~\citep{edgar2002gene} under the accession GSE211692. The raw data was preprocessed by first subtracting the respective background signal from all genes of each sample and then log\textsubscript{2}-transforming the expression values. Subsequently, the whole dataset was normalized by cyclicly applying loess (locally estimated scatterplot smoothing) as implemented in the R package {limma} v3.64.3~\citep{ritchie2015limma}. 
Contrary to the finance dataset, the classes in the genetic dataset are nominal and do not have a clear ordering.

\begin{figure}[ht]
    \centering
    \includegraphics[width=0.9\linewidth]{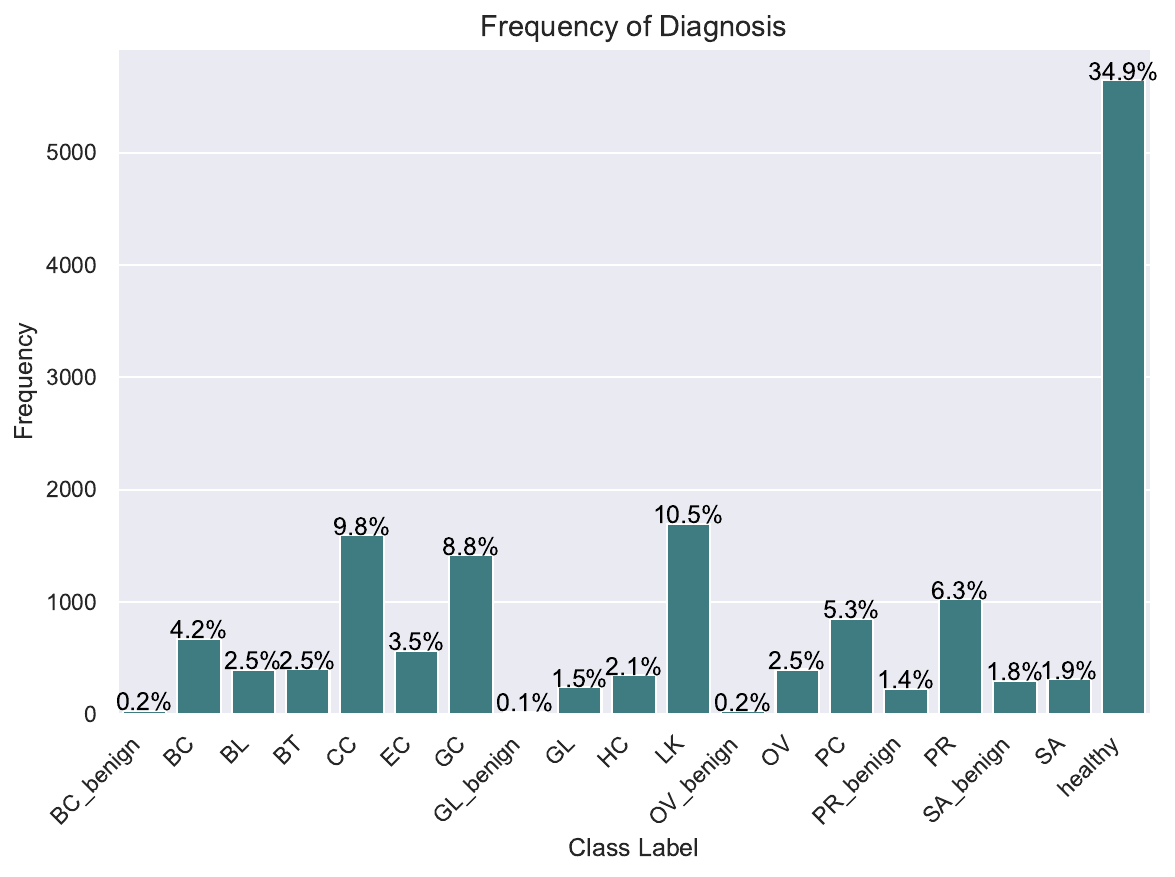}
    \caption{Frequency of diagnosis classes in the genetic dataset}
    \label{fig:class_dist_genetic}
\end{figure}

The class distribution of the data is severely imbalanced, as can be observed in Figure~\ref{fig:class_dist_genetic}. The 5 benign non-cancerous classes are named with the suffix "\_benign", the healthy class is labeled as such, all remaining 13 classes encode the 13 malignant cancer classes. The healthy class makes up 34.9\% of all instances and thus is by far the largest class present, whereas the smallest class (GL\_benign) accounts for only 0.1\% of the dataset. We identify clear class imbalances among the 13 cancer classes, however, the largest differences occur between the healthy class, as majority class, and the 5 benign classes which all are severely underrepresented in the dataset. Nevertheless, it is important to take those samples into account, as their inclusion closely reflects the actual clinical situation. This mirrors a realistic diagnostic use-case in which patients might have other diseases and not solely fall into the more common healthy or cancerous classes.

\begin{figure}[htbp]
\centering
\includegraphics[width=0.9\linewidth]{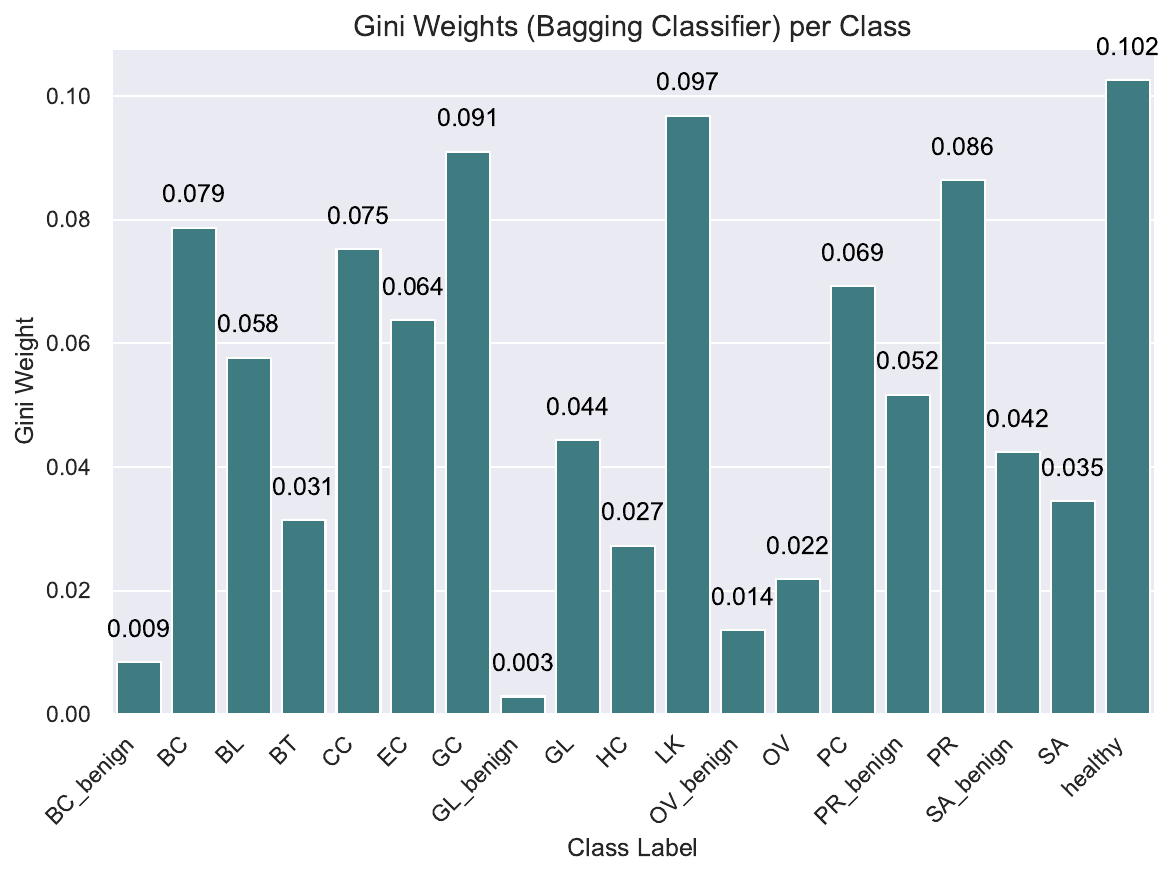}
\caption{Gini weights for a Bagging Classifier showing relative class contributions to the aggregate ROC curve on the genetic dataset.}
\label{fig:weights_genetic}
\end{figure}

We follow the procedure analogous to the finance use-case and evaluate our method using the same machine learning approaches on the clinical data.\\
The weight distribution of Gini weights for a Bagging Classifier run on the genetic dataset is illustrated in Figure~\ref{fig:weights_genetic}. The first thing to note is that the pattern of the Gini weights is much more leveled than the severely imbalanced frequency pattern we observe in Figure~\ref{fig:class_dist_genetic}. Although the differences in class weights are still evident, the values are 
not as extreme as the class distributions.
The healthy class receives the highest weight (0.102), having the highest probability of success. The classes GL\_benign (benign glioma), BC\_benign (benign breast cancer), and OV\_benign (benign ovarian cancer) receive the lowest weights, reflecting their lower probability of success. However, we can clearly observe that the estimated Gini weights per class are not a direct reflection of their frequencies in Figure~\ref{fig:class_dist_genetic}, as the Gini weights are proportionally much larger for these classes relative to their share of the dataset. This shows that the Gini weights simultaneously account for the imbalance of the data as well as the probability of success of the respective classes and thus act like a correction factor.

\begin{figure}[htbp]
\centering
\includegraphics[width=\linewidth]{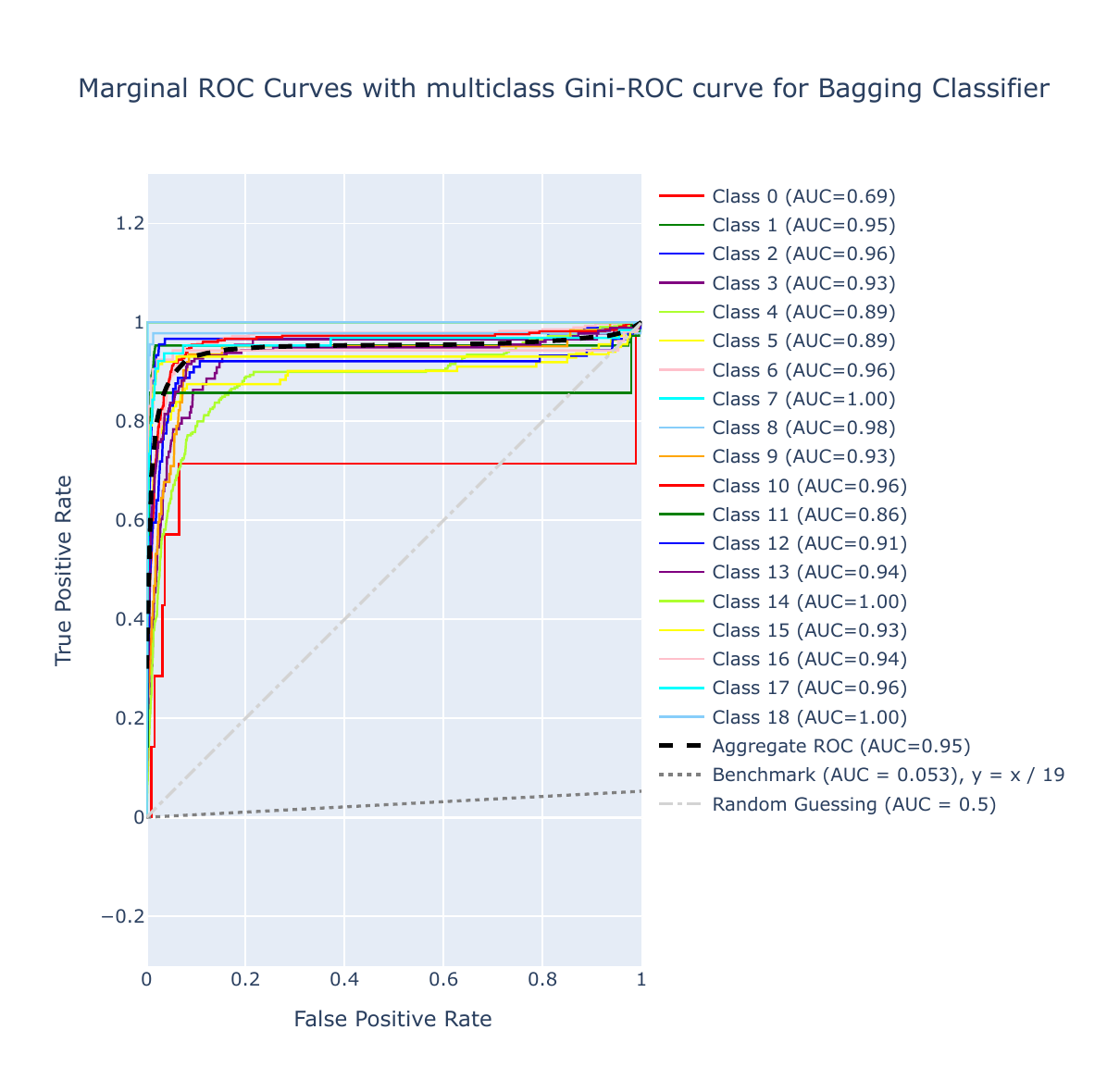}
\caption{Individual ROC curves and multiclass ROC curve for the Bagging Classifier model on the genetic dataset.}
\label{fig:per-class-multiclass-roc-genetic}
\end{figure}

Figure~\ref{fig:per-class-multiclass-roc-genetic} illustrates the comparison of the aggregated multiclass ROC curve with the individual ROC curves of each class. Again, we can observe that the aggregated curve lies between the individual class curves, as expected. In this case, class 0 (BC\_benign) is the hardest to detect, being one of the most imbalanced classes. Class 18, the healthy class, on the other hand, is the easiest to detect, achieving an ideal ROC curve.

The comparison of the Gini AUC with the Micro and Macro AUC in Table~\ref{tab:classifier_metrics_comparison_reduced} shows a similar trend for the genetic dataset as we already discussed for the finance dataset. The Micro and Macro AUCs yield similar values for the genetic dataset across all evaluated models, while the Gini AUCs vary much more. The Bagging Classifier performs best on the health care use case, achieving a Gini AUC of 0.95. Both Micro and Macro AUC are less selective and give higher importance to logistic regression.

\section{CONCLUSION AND FUTURE RESEARCH}
The paper has introduced a novel methodology for constructing multiclass ROC curves, using a multidimensional Gini index, addressing key limitations in existing multiclass evaluation approaches. The methodology delivers a single and interpretable ROC curve that preserves binary analysis properties while capturing multiclass discrimination complexity. 

Traditional multiclass metrics, based on different types of averaging over single-class problems, are not adequate, especially in the presence of highly imbalanced responses. 

Our Gini-based aggregation provides a possible solution that is both scientifically valid, being based on a coherent mathematical framework, and also practically relevant, as it gives higher aggregation weights to the more unbalanced classes.


The methodology can be fruitfully applied to any domain where class imbalance challenges traditional metrics
and regulatory oversight demands an interpretable performance assessment.
It could also be adapted to assess predictions derived from large language models, for which it is difficult to assess accuracy of low frequency classes.

We finally remark that all steps of our proposal can be reproduced using the \texttt{complete\_roc\_analysis} function in the \texttt{multiclass\_roc\_gini} Python package, at:  https://github.com/rosacrg/multiclass-roc-gini.

We used a MacBook Pro, M1 (16GB) running macOS Sequoia Version 15.6.1 to run all experiments.

\subsubsection*{Acknowledgements}
This work is partially supported by the Ministry of Science and Culture of Lower
Saxony through funds from the program zukunft.niedersachsen of the Volkswagen Foundation for the 'CAIMed – Lower Saxony Center for Artificial Intelligence and Causal Methods in Medicine' project (grant no. ZN4257).


  \bibliography{paper}
  \bibliographystyle{plainnat} 

\end{document}